\documentclass[11pt]{article}
\usepackage{graphicx}
\usepackage{amssymb}
\usepackage{amscd}
\usepackage{mathrsfs}
\usepackage{longtable,lscape}
\usepackage{amsthm}
\usepackage{amsfonts}
\usepackage{amsmath}
\usepackage{bbm}
\usepackage{float}
\usepackage{url}
\usepackage{csquotes}
\usepackage{wrapfig}

\usepackage{caption}
\usepackage{lineno}
\usepackage[title]{appendix}
\begin{document}
\title{\bf When Bertlmann wears no socks. Common causes induced by measurements as an explanation for quantum correlations
}
\author{Diederik Aerts and Massimiliano Sassoli de Bianchi\vspace{0.5 cm} \\ 
\normalsize\itshape
Center Leo Apostel for Interdisciplinary Studies, \\ \itshape Brussels Free University, 1050 Brussels, Belgium\vspace{0.5 cm} \\ 
\normalsize
E-Mails: \url{diraerts@vub.ac.be}, \ \url{msassoli@vub.ac.be}
}
\date{}
\maketitle
\begin{abstract} 
\noindent It is well known that correlations produced by common causes in the past cannot violate Bell's inequalities. This was emphasized by Bell in his celebrated example of Bertlmann's socks. However, if common causes are induced by the very measurement process i.e., actualized at each run of a joint measurement, in a way that depends on the type of joint measurement that is being executed (hence, the common causes are contextually actualized), the resulting correlations are able to violate Bell's inequalities, thus providing a simple and general explanation for the origin of quantum correlations. We illustrate this mechanism by revisiting Bertlmann's socks example. In doing so, we also emphasize that Bell's inequalities, in their essence, are about demarcating `non-induced by measurements' (non-contextual) from `induced by measurements' (contextual) common causes, where the latter would operate at a non-spatial level of our physical reality, when the jointly measured entangled entities are microscopic in nature. 
\end{abstract} 
\medskip
{\bf Keywords:} Bell's inequalities; entanglement; no-signaling; contextuality; common causes; extended Bloch representation; hidden-measurement interactions
\vspace{0.5cm}

\section{Introduction}
\label{Introduction}

In a groundbreaking paper, inspired by the analysis of the EPR paradox situation in quantum theory \cite{EinsteinPodolskyRosen1935}, John Bell derived in 1964 mathematical inequalities that are able to test our common sense conception of physical reality, when we are confronted with the phenomenon of quantum entanglement \cite{Bell1964}. For quite some time, his results remained largely ignored by the physics community, but things changed when in the Seventies of last century Alain Aspect proposed a specific experimental scheme using entangled photons and polarizers having their orientations randomly changing in time, credibly testing the violation predicted by quantum mechanics \cite{aspect1976}, and that in the Eighties he performed experiments (with the collaboration of Philippe Grangier, G\'erard Roger and Jean Dalibard) \cite{aspect1981,aspect1982a,aspect1982b,aspect1985} showing for the first time, in a convincing way, the actual violation of Bell's inequalities, later confirmed by numerous experimental groups in the decades to come (see also \cite{Bertlmann1990} for a detailed description of these historical experiments and \cite{hensen-etal2016} for a recent test).

Also in the Eighties, in an article entitled ``Bertlmann's socks and the nature of reality,'' Bell famously mentioned the habit of his colleague Reinhold Bertlmann of always wearing socks of different colors, to point out the difference between quantum and non-quantum correlations, emphasizing that the color-correlations one can observe when meeting Dr.~Bertlmann have nothing to do with those produced by quantum entanglement \cite{Bell1981}.\footnote{The episode was reported by Bertlmann as follows \cite{Bertlmann2015}: ``One day I was sitting in our computer room with my computer cards, when my colleague Gerhard Ecker rushed in, waving a preprint in his hands. He shouted, ``Reinhold, look, now you're famous!'' I could hardly believe my eyes as I read and reread the title of a paper by John, ``Bertlmann's socks and the nature of reality.'' I was totally stunned. As I read the first page, my heart stood still.''} 
In this article, we consider a variation of Bell's story about Bertlmann's socks, to explain that the fundamental demarcation his inequalities provide is between common causes that are induced by the measurement processes (and because of that they are contextual) and common causes that are not induced by the measurement processes (and because of that they are not contextual). 

Our fictional Dr.~Bertlmann (whose habits are slightly different from those of the real Dr.~Bertlmann, as described by Bell) is known to always wear socks of different colors. If one of his two socks is pink, with certainty the other will be observed to be non-pink, hence, a (anti)correlation will be invariably present between the colors of his two socks. The Dr.~Bertlmann of our story also loves pink very much, and always makes sure to start by wearing a pink sock, on the foot he initially pays attention to, then immediately after he wears a non-pink sock, on the other foot. He always takes care to also have a handkerchief in each one of the two front pockets of his trousers, but in this case he likes to always have them of the same color, which fifty percent of the times will be pink and the other fifty percent of the times non-pink.

From time to time, Bertlmann meets his two friends Alice and Bob, who always go around together. When this happens, Alice and Bob are very interested in knowing about Bertlmann's socks and handkerchiefs. More precisely, when Alice meets Bertlmann, depending on her mood, she decides to perform one of the following two measurements, relative to Bertlmann's left side of his body. Measurement $A$ is about the pinkness property of Bertlmann's left handkerchief. To perform it, she simply asks Bertlmann to tell her the color of his left handkerchief. If the answer is ``pink,'' the observation is considered to be successful and the outcome is denoted $A_1$. Otherwise, the outcome is denoted $A_2$. The second measurement that Alice can decide to perform, denoted $A'$, is about the color-correlation between Bertlmann's left handkerchief and left sock. To perform it, she asks Bertlmann to tell her if the handkerchief in his left pocket and his left sock are both pink or both non-pink. If his response is affirmative, the observation is considered to be successful and the outcome is denoted $A'_1$. Otherwise, the outcome is denoted $A'_2$. Bob operates exactly in the same way as Alice, but with respect to Bertlmann's right side of his body. His two measurements are denoted $B$ and $B'$, with outcomes $B_1$ and $B_2$, and $B'_1$ and $B'_2$, respectively. 

So, when Bertlmann meets Alice and Bob, they will jointly ask him the questions relative to the measurements they decide to perform, according to their mood of the moment. This defines four possible joint measurements, which we will denote $AB$, $AB'$, $A'B$ and $A'B'$, where $AB$ is the joint measurement where Alice performs $A$ and Bob performs $B$, $AB'$ is the joint measurement where Alice performs $A$ and Bob performs $B'$, and so on. Assuming for simplicity that Bertlmann has the same propensity to put the pink sock on the left foot or on the right one, it is easy to convince oneself that we have the probabilities: 
\begin{eqnarray}
&&{\cal P}(A_1,B_1)={1\over 2},\quad {\cal P}(A_1,B_2)=0,\quad {\cal P}(A_2,B_1)=0,\quad {\cal P}(A_2,B_2) = {1\over 2},\nonumber\\
&&{\cal P}(A_1,B'_1)={1\over 4},\quad {\cal P}(A_1,B'_2) = {1\over 4},\quad {\cal P}(A_2,B'_1)={1\over 4},\quad {\cal P}(A_2,B'_2)={1\over 4},\nonumber\\
&&{\cal P}(A'_1,B_1)={1\over 4},\quad {\cal P}(A'_1,B_2) = {1\over 4},\quad {\cal P}(A'_2,B_1)={1\over 4},\quad {\cal P}(A'_2,B_2)={1\over 4},\nonumber\\
&&{\cal P}(A'_1,B'_1)=0,\quad {\cal P}(A'_1,B'_2) = {1\over 2},\quad {\cal P}(A'_2,B'_1)={1\over 2},\quad {\cal P}(A'_2,B'_2)=0,
\label{prob-non-contextual}
\end{eqnarray}
which we can use to calculate the expectation values: 
\begin{eqnarray}
&&E(A,B)={\cal P}(A_1,B_1) + {\cal P}(A_2,B_2) - {\cal P}(A_1,B_2)- {\cal P}(A_2,B_1)= 1\nonumber\\
&&E(A,B')={\cal P}(A_1,B'_1) + {\cal P}(A_2,B'_2) - {\cal P}(A_1,B'_2)- {\cal P}(A_2,B'_1)= 0\nonumber\\
&&E(A',B)={\cal P}(A'_1,B_1) + {\cal P}(A'_2,B_2) - {\cal P}(A'_1,B_2)- {\cal P}(A'_2,B_1)=0\nonumber\\
&&E(A',B')={\cal P}(A'_1,B'_1) + {\cal P}(A'_2,B'_2) - {\cal P}(A'_1,B'_2)- {\cal P}(A'_2,B'_1)= -1.
\label{expectations-non-contextual}
\end{eqnarray}

Clearly, the Bell-CHSH inequality \cite{clauser1969}, $|{\rm CHSH}|\leq 2$, with ${\rm CHSH}$ given by the combination $E(A,B)+E(A,B') + E(A',B) - E(A',B')$, or by any other combination obtained by interchanging the roles of $A$ and $A'$ and/or those of $B$ and $B'$, is not violated by the above averages. This because, even though correlations exist between Alice's and Bob's outcomes, they are all the consequence of common causes in the past, which were actualized before Alice and Bob decided to select and perform their measurements. We shall call these past common causes `non-induced by measurements common causes', because they are not actualized during and by the experimental contexts created by Alice and Bob. Note that a straightforward consequence of a common cause not being induced by the measurement process is that Alice's outcome probabilities cannot depend on which measurement Bob decides to jointly perform with her, and vice versa, which means that the so-called no-signaling conditions (also called marginal laws) are automatically obeyed.

\section{When Bertlmann wears no socks}
\label{notwearsocks}

What if the presence of the common causes, responsible for the observed correlations, would be instead induced by the measurements set-ups themselves? To explore this possibility, we now 
consider a different situation.\footnote{Our example is inspired by the macroscopic non-local box presented in \cite{AertsS2005}.} We assume that during the summer period, because of the heat, our fictional Dr.~Bertlmannn decides in the morning not to wear any socks. However, not being sure if he will need them later on during the day, he always keeps a pair of them in his briefcase (a pink one and a non-pink one). On the other hand, he continues to put handkerchiefs of same color in the two pockets of his trousers, as usual. 

This time, when he meets his friends, and Alice asks him to tell if his left handkerchief and his left sock are color-correlated (assuming Alice is performing $A'$ and Bob is performing $B$), since he wears no socks, but is eager not to disappoint his friend and respond to her question, he quickly turns around and puts a sock on his left foot, following his habit of always wearing a pink sock on the foot he initially pays attention to, which in the present situation is the left one, because of Alice's question. Following this action, he can respond to Alice's interrogation, and of course when he puts the pink sock on the left foot, he will also put the remaining non-pink sock on the right foot. The situation where Alice performs $A$ and Bob performs $B'$ is similar, with this time the pink sock ending up on the right foot. On the other hand, when Alice and Bob jointly perform measurements $A'$ and $B'$, Bertlmann has to break the symmetry of the situation by deciding himself on which foot to pay attention first, which will be the one receiving the pink sock. 

Assuming as before that Bertlmann has the same propensity to put the pink sock on the left foot or on the right foot, it is easy to convince oneself that we now have the probabilities: 
\begin{eqnarray}
&&{\cal P}(A_1,B_1)={1\over 2},\quad {\cal P}(A_1,B_2)=0,\quad {\cal P}(A_2,B_1)=0,\quad {\cal P}(A_2,B_2) = {1\over 2},\nonumber\\
&&{\cal P}(A_1,B'_1)={1\over 2},\quad {\cal P}(A_1,B'_2) = 0,\quad {\cal P}(A_2,B'_1)=0,\quad {\cal P}(A_2,B'_2)={1\over 2},\nonumber\\
&&{\cal P}(A'_1,B_1)={1\over 2},\quad {\cal P}(A'_1,B_2) = 0,\quad {\cal P}(A'_2,B_1)=0,\quad {\cal P}(A'_2,B_2)={1\over 2},\nonumber\\
&&{\cal P}(A'_1,B'_1)=0,\quad {\cal P}(A'_1,B'_2) = {1\over 2},\quad {\cal P}(A'_2,B'_1)={1\over 2},\quad {\cal P}(A'_2,B'_2)=0.
\label{prob-contextual}
\end{eqnarray}
Note that (\ref{prob-contextual}) differs from (\ref{prob-non-contextual}) only for what concerns the probabilities relative to the situations where Alice and Bob perform different typologies of measurements. To understand why, consider that there are only two possibilities prior to the execution of the joint measurements: Bertlmann has pink handkerchiefs in his two pockets, or non-pink ones. Since he wears no socks, in the joint measurements $AB'$ the pink sock will go to his right foot and the non-pink one to his left foot. Hence, we have two possible outcomes: one such that the handkerchiefs are pink and the right sock is also pink, which corresponds to outcome $(A_1,B'_1)$, the other such that the handkerchiefs are non-pink and the right sock is pink, i.e., not of the same color of the handkerchiefs, which corresponds to outcome $(A_2,B'_2)$. Hence, ${\cal P}(A_1,B'_1)={1\over 2}$ and ${\cal P}(A_2,B'_2)={1\over 2}$, and we can reason in the same way for the other probabilities in (\ref{prob-contextual}). 

The expectation values are now: 
\begin{eqnarray}
&&E(A,B)={\cal P}(A_1,B_1) + {\cal P}(A_2,B_2) - {\cal P}(A_1,B_2)- {\cal P}(A_2,B_1)= 1\nonumber\\
&&E(A,B')={\cal P}(A_1,B'_1) + {\cal P}(A_2,B'_2) - {\cal P}(A_1,B'_2)- {\cal P}(A_2,B'_1)= 1\nonumber\\
&&E(A',B)={\cal P}(A'_1,B_1) + {\cal P}(A'_2,B_2) - {\cal P}(A'_1,B_2)- {\cal P}(A'_2,B_1)=1\nonumber\\
&&E(A',B')={\cal P}(A'_1,B'_1) + {\cal P}(A'_2,B'_2) - {\cal P}(A'_1,B'_2)- {\cal P}(A'_2,B'_1)= -1,
\label{expectations-contextual}
\end{eqnarray}
which means that the Bell-CHSH inequality is violated up to its algebraic maximum, as is clear that $E(A,B)+E(A,B') + E(A',B) - E(A',B')=4$. 

Note that the no-signaling conditions are also satisfied. Indeed, ${\cal P}_B(A_1)\equiv {\cal P}(A_1,B_1)+{\cal P}(A_1,B_2)={1\over 2}+0={1\over 2}$ and ${\cal P}_{B'}(A_1)\equiv {\cal P}(A_1,B'_1)+{\cal P}(A_1,B'_2)={1\over 2}+0={1\over 2}$. Also, ${\cal P}_B(A_2)\equiv {\cal P}(A_2,B_1)+{\cal P}(A_2,B_2)=0+{1\over 2}={1\over 2}$ and ${\cal P}_{B'}(A_2)\equiv {\cal P}(A_2,B'_1)+{\cal P}(A_2,B'_2)=0+{1\over 2}={1\over 2}$. In other words, ${\cal P}_B(A_i)={\cal P}_{B'}(A_i)$, $i=1,2$, and it is easy to check that we also have, with obvious notation, ${\cal P}_B(A'_i)={\cal P}_{B'}(A'_i)$, ${\cal P}_A(B_i)={\cal P}_{A'}(B_i)$ and ${\cal P}_A(B'_i)={\cal P}_{A'}(B'_i)$, $i=1,2$. In other words, Alice and Bob, with their joint measurements, can create correlations but cannot use them to communicate.

\section{Induced by measurements common causes}

The above example of `Bertlmann wearing no socks' shows that the idea that quantum correlations would result from `induced by measurements'  
common causes' is a perfectly licit one. However, despite its simplicity and generality, it has surprisingly not attracted significant interest in the scientific community so far (although the idea was first presented, albeit with a slightly different terminology, since the eighties of the last century \cite{Aerts1984}). Note that in this article our focus is uniquely on the `Bertlmann wearing no socks' situation, because of its affinity with Bell's historical example, but it is possible and easy to replace Dr.~Bertlmann by a passive physical entity, like an elastic band, with the outcomes being for instance created by Alice and Bob when they jointly pull its two ends, causing it to break at some unpredictable point, so that the lengths of the obtained fragments will necessarily be correlated; see \cite{AertsS2005} for the details. 

Note also that the very quantum formalism already suggests that when a system is not in an eigenstate of the measurement being executed, the outcome is literally created (i.e., actualized) by the latter, and when a joint measurement is executed, joint outcomes will consequently be actualized; and if they are correlated, it is natural to assume that the joint measurement also actualized the common causes that are at the origin of the correlations. Also, if the common causes are actualized only when the joint measurements are executed, there will be no `unique set of common causes' characterizing all the correlations of the four joint measurements $AB$, $AB'$, $A'B$ and $A'B'$, which is the reason why the Bell-CHSH inequality can be violated (as we will better explain in the next section).

Quantum micro-physical entities in entangled states can be described similarly to the situation of Bertlmann wearing no socks. For example, a system formed by two electrons in a singlet state can be conceptualized as a situation where the two electrons ``wear no spin,'' i.e., a situation where the two-entities haven't yet actualized specific spin directions, which therefore remain potential until the meeting with Alice's and Bob's Stern-Gerlach apparatuses, forcing them to acquire one, in the same way Alice and Bob, when they meet Bertlmann, also ``force'' him to wear socks of specific colors. This can be more clearly seen by also observing that a singlet state $|s\rangle \propto |+\rangle_{\hat{u}} |-\rangle_{\hat{u}} - |-\rangle_{\hat{u}} |+\rangle_{\hat{u}}$, where $|+\rangle_{\hat{u}}$ and $|-\rangle_{\hat{u}}$ represent the ``up'' and ``down'' eigenstates of the one-entity spin observables along the $\hat{u}$-direction, is a rotationally invariant vector. This means that there is nothing particular about the direction $\hat{u}$ that is used to explicitly write it, in terms of ``up'' and ``down'' components: one can equivalently choose a different direction $\hat{w}\neq \hat{u}$ and still find, after some algebraic calculation, that $|s\rangle\propto |+\rangle_{\hat{w}} |-\rangle_{\hat{w}} - |-\rangle_{\hat{w}} |+\rangle_{\hat{w}}$. In other words, the zero-spin state $|s\rangle$ resides in a one-dimensional subspace of the two-entity Hilbert space (obtained from the antisymmetric tensor product of the two one-entity state spaces) which is invariant under the action of the 3D rotation group. Hence, it is to be interpreted as a ``no-spin state,'' where spins are genuinely potential, thus non-spatial. In our Bertlmann's example (which could receive an explicit quantum mathematical representation in complex Hilbert space, following the method presented in \cite{as2014}), we can easily see how the the choice of the joint measurement to be performed (the experimental context) 
affects the outcomes' correlations. Measurement $A'B'$ is an indeterministic context as regards the foot receiving the pink sock. Indeed, because of the questions jointly addressed by Alice and Bob, Bertlmann has to actualize a thought, or sensation, about one of his two feet, on which he will bring his attention. Such thought, or sensation, in combination with Bertlmann's habit of wearing first the pink sock, is an example of a `induced by measurements (contextual) common cause', 
i.e., of a common cause which will not be the same when a different joint measurement is executed. In fact, measurements $A'B$ and $AB'$ are deterministic contexts as regards the foot that will receive the pink sock: $A'B$ always produces a left pink sock and $AB'$ always produces a right pink sock. So, the common cause is different from that of measurement $A'B'$ -- hence contextual -- as it does not involve anymore a decision-making from Bertlmann, whose initial attention towards one of his feet is now externally triggered by Alice, in measurement $A'B$, or by Bob, in measurement $AB'$. 

The fundamental distinction between `non-induced by measurements' and `induced by measurements'  common causes was emphasized by one of us already in 1990, however not directly referring to the causes of the correlations, but to the correlations themselves. More precisely, correlations resulting from `non-induced by measurements' (non-contextual) common causes were referred to as `correlations of the first kind', whereas correlations originating from `induced by measurements' (contextual) common causes, actualized during the measurement processes, were referred to as `correlations of the second kind' \cite{Aerts1990}. Among the reasons that led us to write this article, there is also that of proposing this new designation, which we hope will be able to more efficiently capture the attention of physicists on this simple and natural explanation for quantum correlations.

\section{Bell's locality assumption}

When two outcomes are correlated, their joint probability will not factorize into a product of probabilities for the individual outcomes, and this independently of the fact that the correlations are of the first or of the second kind, i.e., resulting from `non-induced by measurements' or `induced by measurements' common causes, respectively. 

Let us consider first the ``standard'' situation of Bertlmann leaving home with socks already on his feet. According to (\ref{prob-non-contextual}), we have ${\cal P}(A_1)={\cal P}(A_1,B_1)+{\cal P}(A_1,B_2)={\cal P}(A_1,B'_1)+{\cal P}(A_1,B'_2)={1\over 2}$ and ${\cal P}(B_1)={\cal P}(A_1,B_1)+{\cal P}(A_2,B_1)={\cal P}(A'_1,B_1)+{\cal P}(A'_2,B_1)={1\over 2}$. On the other hand, ${\cal P}(A_1,B_1)={1\over 2}$, hence ${\cal P}(A_1,B_1)\neq {\cal P}(A_1){\cal P}(B_1)$. 

As emphasized by Bell, for example in \cite{Bell1981}, it is however reasonable to assume that if the common causes of the correlations are known and can be held fixed, the probabilities associated with possible residual fluctuations will factorize. This is what is usually called `Bell's locality assumption'. Let us denote by $\lambda_i$, $i=1,2,3,4$, the common causes at the origin of the correlations of the first kind (\ref{prob-non-contextual}). These can be identified with the four possible states of mind of Bertlmann when wearing his socks and handkerchiefs. More precisely, $\lambda_1$ describes the state of mind such that Bertlmann decides to wear pink handkerchiefs and a pink left sock (and therefore a non-pink right sock); $\lambda_2$ describes the state of mind such that Bertlmann decides to wear pink handkerchiefs and a non-pink left sock; $\lambda_3$ describes the state of mind such that Bertlmann decides to wear non-pink handkerchiefs and a non-pink left sock; $\lambda_4$ describes the state of mind such that Bertlmann decides to wear non-pink handkerchiefs and a pink left sock.

Consider, as an example, the situation where we know that the state of mind of Bertlmann is $\lambda_1$. Then the probabilities become: 
\begin{eqnarray}
&&{\cal P}(A_1,B_1;\lambda_1)=1,\quad {\cal P}(A_1,B_2;\lambda_1)=0,\quad {\cal P}(A_2,B_1;\lambda_1)=0,\quad {\cal P}(A_2,B_2;\lambda_1) = 0,\nonumber\\
&&{\cal P}(A_1,B'_1;\lambda_1)=0,\quad {\cal P}(A_1,B'_2;\lambda_1) = 1,\quad {\cal P}(A_2,B'_1;\lambda_1)=0,\quad {\cal P}(A_2,B'_2;\lambda_1)=0,\nonumber\\
&&{\cal P}(A'_1,B_1;\lambda_1)=1,\quad {\cal P}(A'_1,B_2;\lambda_1) = 0,\quad {\cal P}(A'_2,B_1;\lambda_1)=0,\quad {\cal P}(A'_2,B_2;\lambda_1)=0,\nonumber\\
&&{\cal P}(A'_1,B'_1;\lambda_1)=0,\quad {\cal P}(A'_1,B'_2;\lambda_1) = 1,\quad {\cal P}(A'_2,B'_1;\lambda_1)=0,\quad {\cal P}(A'_2,B'_2;\lambda_1)=0,
\label{prob-non-contextual-fixed cause}
\end{eqnarray}
and we also have for the marginals: 
\begin{eqnarray}
&&{\cal P}(A_1;\lambda_1)={\cal P}(B_1;\lambda_1)={\cal P}(B'_2;\lambda_1)={\cal P}(A'_1;\lambda_1)=1,\nonumber\\
&&{\cal P}(A_2;\lambda_1)={\cal P}(B_2;\lambda_1)={\cal P}(B'_1;\lambda_1)={\cal P}(A'_2;\lambda_1)=0,
\label{marginal-non-contextual-fixed cause}
\end{eqnarray}
which means that, say for joint measurement $AB'$, we have the factorization: 
\begin{equation}
{\cal P}(A_i,B'_j;\lambda_1)= {\cal P}(A_i;\lambda_1){\cal P}(B'_j;\lambda_1),\quad i,j=1,2,
\label{factorization-non-contextual-fixed cause}
\end{equation}
and the same holds true for the other joint measurements, when the states of mind $\lambda_2$, $\lambda_3$ and $\lambda_4$ are fixed. Also, with the assumption (here just for simplicity) that Bertlmann's different states of mind are equiprobable, it is easy to check that we have the equalities:
\begin{equation}
{\cal P}(A_i,B'_j)={1\over 4}\sum_{k=1}^4 {\cal P}(A_i,B'_j;\lambda_k)= {1\over 4}\sum_{k=1}^4 {\cal P}(A_i;\lambda_k){\cal P}(B'_j;\lambda_k),
\label{average-non-contextual-fixed cause}
\end{equation}
for all $i,j=1,2$, and the same holds for the other joint measurements. 

It immediately follows from (\ref{average-non-contextual-fixed cause}) that the Bell-CHSH inequality will be satisfied. This is a standard derivation, but we reproduce it below for facilitating our subsequent discussion. Using (\ref{average-non-contextual-fixed cause}), we can write: 
\begin{eqnarray}
E(A,B')&=&{\cal P}(A_1,B'_1) + {\cal P}(A_2,B'_2) - {\cal P}(A_1,B'_2)- {\cal P}(A_2,B'_1)\nonumber\\
&=&{1\over 4}\sum_{k=1}^4 [{\cal P}(A_1,B'_1;\lambda_k)+ {\cal P}(A_2,B'_2;\lambda_k)-{\cal P}(A_1,B'_2;\lambda_k)- {\cal P}(A_2,B'_1;\lambda_k)]\nonumber\\
&=&{1\over 4}\sum_{k=1}^4 [{\cal P}(A_1;\lambda_k)- {\cal P}(A_2;\lambda_k)] [{\cal P}(B'_1;\lambda_k) -{\cal P}(B'_2;\lambda_k)]\nonumber\\
&=& {1\over 4}\sum_{k=1}^4 E(A;\lambda_k)E(B';\lambda_k),
\label{average}
\end{eqnarray}
and same for the averages of the other joint measurements. Observing that we can write:
\begin{eqnarray}
&&E(A,B)\pm E(A,B')={1\over 4}\sum_{k=1}^4 E(A;\lambda_k)[E(B;\lambda_k)\pm E(B';\lambda_k)],\nonumber\\
&&E(A',B)\mp E(A',B')={1\over 4}\sum_{k=1}^4 E(A';\lambda_k)[E(B;\lambda_k)\mp E(B';\lambda_k)],
\label{average-difference}
\end{eqnarray}
and that by definition $|E(A;\lambda_k)|, |E(A';\lambda_k)|, |E(B;\lambda_k)|, |E(B';\lambda_k)|\leq 1$, we have: 
\begin{eqnarray}
&&|E(A,B)\pm E(A,B')|\leq {1\over 4}\sum_{k=1}^4 |E(B;\lambda_k)\pm E(B';\lambda_k)|,\nonumber\\
&&|E(A',B)\mp E(A',B')|\leq{1\over 4}\sum_{k=1}^4 |E(B;\lambda_k)\mp E(B';\lambda_k)|.
\label{average-difference-module}
\end{eqnarray}
Finally, observing that we also have $|E(B;\lambda_k)\pm E(B';\lambda_k)|+|E(B;\lambda_k)\mp E(B';\lambda_k)|\leq 2$, we obtain: 
\begin{eqnarray}
&&|E(A,B)+E(A,B') + E(A',B) - E(A',B')|\leq |E(A,B)+E(A,B')|+|E(A',B) - E(A',B')|\nonumber\\
&&\quad \leq {1\over 4}\sum_{k=1}^4 \left[|E(B;\lambda_k)\pm E(B';\lambda_k)|+|E(B;\lambda_k)\mp E(B';\lambda_k)|\right] \leq 2.
\label{CHSH-inequality}
\end{eqnarray}

In other words, although common causes that are not induced by measurements 
do not imply that the latter, when jointly performed by Alice and Bob, would be separate, in the sense that, say for joint measurement $AB'$, we would have ${\cal P}(A_i,B'_j)={\cal P}(A_i){\cal P}(B'_j)$, since we have still  the validity of the average (\ref{average-non-contextual-fixed cause}), they nevertheless imply that the Bell-CHSH inequality is satisfied. 

Let us now consider the situation of Bertlmann wearing no socks, when he meets Alice and Bob, i.e., the situation where the common causes are induced by the measurements. 
Why in this case the Bell-CHSH inequality can be violated? This is so because we cannot anymore write the expectation values of the joint measurements as in (\ref{average}), i.e., as sums of products of the expectation values for the individual measurements, with the sums running over the same common causes. The reason we cannot do this is that now the common causes are not the same for the different joint measurements, i.e., they are contextual. Let us show this in some detail. 

For joint measurement $AB$, where Alice and Bob only ask about the color of Bertlmann's handkerchiefs, there are only two possible causes at the origin of the observed correlations. One corresponds to the state of mind of Bertlmann when he decides to wear pink handkerchiefs, let us denote it $\mu_1$, the other corresponds to the state of mind of Bertlmann when he decides to wear non-pink handkerchiefs, let us denote it $\mu_2$. 

Consider joint measurement $A'B$. There is now a combination of the causes $\mu_1$ and $\mu_2$ with the event of meeting Alice asking about the color-correlation between the left sock and the left handkerchief, and also jointly meeting Bob only asking about the color of the right handkerchief. This triggers a deterministic action from Bertlmann, wearing a pink sock on his left foot (and consequently a non-pink sock on the right foot). So, we have again two possible common causes at the origin of the observed outcomes, let us call them $\nu_1$ and $\nu_2$, which are clearly different from those of joint measurement $AB$, with $\nu_1$ giving rise to outcome $(A'_1,B_1)$ and $\nu_2$ giving rise to outcome $(A'_2,B_2)$. 

For joint measurement $AB'$, reasoning in a similar way, we find that we have again two possible common causes at the origin of the observed outcomes, which we will denote $\sigma_1$ and $\sigma_2$, with the former giving rise to outcome $(A_1,B'_1)$ and the latter to outcome $(A_2,B'_2)$. 

Finally, for joint measurement $A'B'$, the situation is similar to that described in the previous section, with the four common causes $\lambda_1$, $\lambda_2$, $\lambda_3$ and $\lambda_4$. So, we can now write the following four uniform averages of products of probabilities ($i,j=1,2$):
\begin{eqnarray}
&&{\cal P}(A_i,B_j)={1\over 2}\sum_{k=1}^2 {\cal P}(A_i;\mu_k){\cal P}(B_j;\mu_k),\quad {\cal P}(A'_i,B_j)= {1\over 2}\sum_{k=1}^2 {\cal P}(A'_i;\nu_k){\cal P}(B_j;\nu_k),\nonumber\\
&& {\cal P}(A_i,B'_j)={1\over 2}\sum_{k=1}^2 {\cal P}(A_i;\sigma_k){\cal P}(B'_j;\sigma_k),\quad {\cal P}(A'_i,B'_j)={1\over 4}\sum_{k=1}^4 {\cal P}(A'_i;\lambda_k){\cal P}(B'_j;\lambda_k).
\label{average-contextual-fixed-contextual-cause}
\end{eqnarray}
The above allows us to still write for the expectation values: 
\begin{eqnarray}
&&E(A,B)={1\over 2}\sum_{k=1}^2 E(A;\mu_k)E(B;\mu_k),\quad E(A',B)= {1\over 2}\sum_{k=1}^2 E(A';\nu_k)E(B;\nu_k),\nonumber\\
&& E(A,B')={1\over 2}\sum_{k=1}^2 E(A;\sigma_k)E(B';\sigma_k),\quad E(A',B')={1\over 4}\sum_{k=1}^4 E(A';\lambda_k)E(B';\lambda_k).
\label{average-contextual}
\end{eqnarray}
However, when considering the sums and differences in (\ref{average-difference}), we are now unable to factorize the different expressions, as we did in the right hand sides of (\ref{average-difference}), as is clear that the averages are not anymore functions of the same variables for the common causes at the origin of the correlated outcomes. Hence, the Bell-CHSH inequality cannot be proven anymore and can in principle be violated, even up to its maximum value, as it is the case in our example.

\section{Hidden-measurement interactions}

Generally speaking, a measurement is a process involving the interaction of an entity (which is the entity to be measured) with an experimental context, usually described as another entity, called the measuring apparatus. One also needs to specify the operations to be performed in order to correctly obtain such interaction and the rule to be used to interpret its effects, in terms of possible outcomes of the measurement. 

Quantum measurements are typically associated with indeterministic experimental contexts, corresponding to situations such that even when there is a full knowledge of the pre-measurement state, there is a maximum and irreducible lack of knowledge about the interaction with the measurement apparatus. To put it another way, in a quantum measurement the causes producing the effects of the different possible outcomes are induced by the very measurement process (and in that sense are contextual), i.e., actualized in an unpredictable way, as in a (weighted) symmetry breaking process, at each run of the measurement \cite{AertsSassoli2017}. 

This mechanism, based on ``contextual causes,'' which is likely at the origin of quantum indeterminism, becomes particularly evident when the standard formalism is extended in what was called the `extended Bloch representation (EBR)' of quantum mechanics \cite{AertsSassolideBianchi2014,AertsSassoli2016}. It is not the scope of this article to enter into the details of the EBR, which provides a natural completion of the quantum formalism. Let us simply mention here that it uses a generalized Bloch sphere representation in which it is possible to describe not only the states of a measured entity, but also the causes that can contextually produce the different outcomes, so much so that in this approach the Born rule can be derived in a non-circular way, instead of being just postulated. 

In the EBR formalism, the `non-induced by measurements' causes  at the origin of the different possible outcomes are called `hidden-measurement interactions'. However, the term ``interaction'' should not be understood here in the sense of a `fundamental force' described by a potential term in the system's Lagrangian, but as the cause of a symmetry breaking actualizing a potential property of the system, resulting from bringing the measured system in contact with the measuring apparatus (in our example, resulting from Bertlmann meeting with Alice and Bob).\footnote{The term ``hidden,'' in ``hidden-measurement interactions,'' is to be understood in the sense of ``inaccessible,'' i.e., in the sense of an aspect of the measurement process that the experimenter cannot take control of, without altering the experiment in such a way that it would not correspond anymore to the observation of the same physical property.}

Quantum joint measurements are just a specific type of measurements performed on bipartite entities, where the operations to be carried out on the two components forming the entity, traditionally described as Alice's and Bob's measurements, can be distinguished and are performed in a coincident way, with the resulting effects being described as couples of outcomes. When the entity is in an entangled state, these couples of outcomes will be correlated in such a way that Bell's inequalities can be violated. From the perspective of the EBR, this is what one would expect to happen. Indeed, the couples of outcomes produced by Alice's and Bob's joint measurements are precisely described in this approach as the effects of common causes actualized at each run of the experiment, in a way that depends on which measurements Alice and Bob have decided to jointly execute, i.e., in a contextual way. In other words, the situation described in our above simplified example of Dr.~Bertlmann (when he does not wear socks) does capture the essence of a mechanism which can be described in all generality within the mathematical formalism of the EBR of quantum mechanics.

\section{Final remarks}

Our analysis shows that the fundamental ingredient for obtaining a violation of the Bell-CHSH inequality is that the common causes at the origin of the observed correlations when, say, $AB$ is performed, are different from those at the origin of the observed correlation when, say, $A'B$ is performed, as evidenced in (\ref{average-contextual-fixed-contextual-cause}). In the quantum formalism, this manifests in the fact that the observables associated with $AB$ and $A'B$ do not commute, and when these observables are written as tensor products, this of course reduces to the non-commutability of $A$ and $A'$. 

However, as the `Dr.~Bertlmann wearing no socks' experimental situation demonstrates, $A$ and $A'$ (and similarly $B$ and $B'$) can very well be compatible (i.e., jointly executable) measurements and one can still have a violation of the Bell-CHSH inequality, also a maximal one for that matter. So, the incompatibility responsible for the violation of the Bell-CHSH inequality truly lies at the non-local level of the overall joint measurements, which are characterized by different sets of hidden-measurement interactions, i.e., by different common causes, actualized in a contextual way. 

Bell's inequalities are in that sense to be understood as providing a fundamental demarcation between the following two situations. When they are obeyed, it means that the observed correlations are the result of common causes that are contextually induced by the measurements themselves; when they are violated, it means that the common causes are not induced by the measurements, or at least not in a way that is different for each measurement. 
Also, this mechanism of the measurements being at the origin of the observed correlations, is not a prerogative of the micro-physical systems only, as the many examples of macroscopic entities violating Bell's inequalities analyzed during the years have clearly shown (see \cite{AertsEtAl2019} and the references cited therein). 

The important difference, when Bell's inequalities are violated by classical macro-physical entities, like Dr.~Bertlmann's socks and handkerchiefs, compared to when they are violated by quantum micro-physical ones, like the entangled spins of two electrons, is that the connections out of which the common causes are created for the macro-entities are spatial elements of reality, whereas for the micro-entities they are non-spatial. Note that our fictional Dr.~Bertlmann character can also be interpreted as a bipartite entity, as we can easily distinguish his left foot and left pocket from his right foot and right pocket. These left and right physical aspects, although spatially separated, remain intimately connected at a more abstract level, through Bertlmann's habits of always wearing socks of different colors and handkerchiefs of the same color.\footnote{In that respect, our example is somehow in between spatiality and non-spatiality. Indeed, where are Bertlmann's habits located? Are they in space? These are challenging questions that will receive very different answers depending on the person to whom they are asked.}

The same is true when considering micro-entities, like two entangled electrons in a singlet state. Even if the two electrons, once they have traveled far away from the source, can only be detected in widely separated spatial regions, they nevertheless remain connected at a more abstract level, thus manifesting a genuine non-spatial nature, allowing them to still behave as a whole entity. And when Alice and Bob jointly act on such an interconnected bipartite entity, correlations will be contextually created out of its whole structure. In other words, the common causes at the origin of the correlations have to be understood not only as being induced by measurements, and therefore genuinely contextual, but also genuinely non-spatial. 

In our `Dr.~Bertlmann not wearing socks' experimental situation, the so-called no-signaling conditions (also called marginal laws) are satisfied, in accordance with what is predicted by the quantum formalism, when the joint measurements are assumed to be correctly described by tensor product observables. In other words, the statistics of outcomes that Alice obtains does not depend on the measurements Bob is jointly performing, and vice versa, so that Alice and Bob cannot use their joint measurements to communicate \cite{GhirardiEtAl1980,Ballentine1987}. However, it is easy to define measurements such that the marginals laws will also be violated. This is the case for instance for the historical `vessels of water model' \cite{Aerts1982}, where joint measurements on a system formed by two vessels connected through a tube and containing a certain amount of transparent water were considered, and for similar models that have been analyzed during the years \cite{AertsEtAl2019}. 
In the `Dr.~Bertlmann not wearing socks' situation, if instead of $A'$ we consider a measurement $A''$, consisting in asking Bertlmann to tell the color of his left sock, with $A''_1$ corresponding to the ``pink'' answer and $A''_2$ to the ``non-pink'' answer, and similarly for $B''$, for the right sock, then it is easy to check that the four joint measurements $AB$, $AB''$, $A''B$ and $A''B''$ will also maximally violate the Bell-CHSH inequality, but this time will also violate the no-signaling conditions.

The reason we have presented in this article a situation not violating the marginal laws, is that we wanted to emphasize that the `induced by measurements' mechanism is fully compatible with the no-signaling requirement, but it is important to also emphasize that it is general enough to also allow for the description of situations where the marginal laws are not necessarily obeyed. Now, in many experiments the latter have been observed to be violated \cite{AdenierKhrennikov2007,DeRaedt2012,DeRaedt2013,AdenierKhrennikov2016,Bednorz2017,Kupczynski2017}, although it is still not clear what the correct interpretation of these violations should be. Note however that when an entangled bipartite system is understood as forming an undivided non-spatial whole, and joint measurements are seen as processes contextually actualizing common causes producing the correlations, in principle the latter can violate not only the Bell-CHSH inequality but also the marginal laws. This does not mean, as is usually believed, that a superluminal communication would for this be possible. Indeed, if space-like separated correlations originating from a common cause in the past are not in violation of relativity theory (as `Bertlmann's socks' example illustrates), then the same will also be true if the common cause is induced by the very measurement. This means that experimental situations where `induced by measurements common causes' also give rise to a violation of the marginal laws cannot be used to achieve any (statistical) faster-than-light communication, as no genuine faster-than-light signal traveling in space can be initiated and controlled by either Alice or Bob. For more details on this specific aspect, we refer the interested reader to \cite{AertsEtAl2019}.

We observe that it is quite well accepted among physicists that quantum mechanics cannot be embedded in a `locally causal theory', i.e., a theory such that, to quote Bell \cite{Bell2004}: ``the direct causes (and effects) of events are near by, and even the indirect causes (and effects) are no further away than permitted by the velocity of light.'' In this article, we have emphasized that the reason for the irreducible non-locality of the theory would not be the existence of some unknown signals propagating ``quantum information'' in space at some superluminal speeds (which should be of at least seven orders of magnitude larger than the speed of light, according to \cite{ZbindenEtAl2001}), but a mechanism of actualization of potential 
common causes, during the execution of joint measurements, explaining why space-like separated correlations can be easily created violating Bell's inequalities.

We also mention that when we introduce `induced by measurements common causes', hence contextual common causes, as an explanation for the violation of Bell's inequalities, we do not want to merely highlight a well-known result, namely that certain quantum effects are incompatible with non-contextual hidden variable models. The new insight we want to put forward here is that the `induced by measurements common causes' are physically real and not just the result of an \emph{ad hoc} mathematical construction. Indeed, in our `Bertlmann wears no socks' example, the `induced by measurements common causes' are easy to point out, hence are not some mysterious ``hidden'' variables. The same is true in many other experiments with macroscopic objects, like an elastic band that is pulled from its two ends until it breaks into two fragments of correlated lengths \cite{AertsS2005,Sassoli2013}, or a volume of water that is extracted using two siphons, thus producing two correlated quantities of water, in two separate reference containers \cite{Aerts1982}, or two dice connected via a rigid rod that are jointly rolled and thus produce correlated pairs of ``upper faces'' \cite{Sassoli2013b}, or even in experiments with abstract entities, like specific conceptual combinations that are brought into more concrete states in well-designed psychological experiments (see \cite{AertsEtAl2019} and the references cited therein). 

This is just to emphasize that experimental contexts designed in such a way that the joint measurements will induce the common causes actualizing the correlations, will naturally violate Bell's inequalities, regardless of whether these experiments are performed on entities that are quantum or classical, microscopic or macroscopic, abstract or concrete. Of course, when performing experiments on micro-physical entities, we cannot access the level where the `induced by measurements common causes' are actualized, as we can do instead with the `Bertlmann wears no socks' example and the other above-mentioned examples. However, considering that two micro-physical entities in entangled states do form a whole that gets separated by joint measurements, similarly to how a whole elastic gets separated into two fragments when pulled from its ends, an explanation of the origin of quantum correlations in terms of `common causes induced by measurements'  is a very natural and general one, applicable to all sorts of systems. 

When a rock explodes in two pieces flying apart, many properties of the two pieces of rock will be correlated. Similarly, when two pieces of a same playing card are hidden in two envelopes and sent to distant places, where they are opened, the outcomes will be correlated (for instance, their color will be the same, either red or black). Such examples are typically given as situations where Bell's inequalities are not violated, in contrast to the more mysterious character of a quantum violation. But as we explained in this article, the reason for this difference is that the measurements considered on the `flying apart pieces of rocks', or the `remote opening of the envelopes', are not as such `inducing' the common causes at the origin of the correlations, as is the case for the `breaking of the elastic'. Instead, they only `observe already existing correlations and register them in a passive way'. In other words, in our highlighting that Bell's inequalities are naturally violated by `common causes induced by measurements', in relation to both microscopic and macroscopic entities, the fact that the measurements produce the correlations, instead of merely observing them, is absolutely crucial.

\end{document}